\documentclass[12pt]{article}

\usepackage[utf8]{inputenc}
\usepackage{amsmath}
\usepackage{amsthm}
\usepackage{amsfonts}
\usepackage{mathtools}
\usepackage{amssymb}
\usepackage{comment}
\usepackage{dsfont}
\usepackage{stackengine}
 \usepackage{booktabs}
 \usepackage{enumitem}
  \usepackage{pdfpages}

\usepackage[hidelinks]{hyperref}

\usepackage{tikz}
\usepackage{dblfloatfix} 
\usepackage[labelfont=bf]{caption}
\usepackage{subcaption}
\usepackage{tkz-tab}
\usepackage{dsfont}
\usepackage{wrapfig}
\usepackage{wasysym}
\usepackage{enumitem}
\usepackage{adjustbox}
\usepackage{ragged2e}
\usepackage{longtable}
\usepackage{changepage}
\usepackage{setspace}
\usepackage{hhline}
\usepackage{multicol}
\usepackage{tabto}
\usepackage{float}
\usepackage{multirow}
\usepackage{makecell}
\usepackage{fancyhdr}
\usepackage{mathdots}

\usepackage{times}

\usepackage{algorithm}
\usepackage{algpseudocode}%

\newenvironment{sciabstract}{%
\begin{quote} \bf}
{\end{quote}}

\usepackage{dsfont}
\usepackage{amsfonts}
\usepackage{amsthm}
\usepackage{amsmath}
\usepackage{amscd}
\usepackage{amssymb}
\usepackage{enumitem}

\usepackage{float}

\usepackage{caption}
\usepackage{subcaption}

\usepackage{comment}

\newtheorem{Th}{Theorem}

\theoremstyle{definition}

\newtheorem{?}[Th]{Problem}

\usepackage{mathtools}
\mathtoolsset{showonlyrefs}

\usepackage{wasysym}
\usepackage{yhmath}

\title{\LARGE{Long ties accelerate noisy threshold-based contagions}}

\author{Dean Eckles$^{a,\ast}$, Elchanan Mossel$^{b}$, M. Amin Rahimian$^{c,\ast}$ and Subhabrata Sen$^d$\\
\normalsize{$^a$~Sloan School of Management, Massachusetts Institute of Technology \vspace{-1mm}}\\
\normalsize{$^b$~Department of Mathematics, Massachusetts Institute of Technology \vspace{-1mm}}\\
\normalsize{$^c$~Department of Industrial Engineering, University of Pittsburgh \vspace{-1mm}}\\
\normalsize{$^d$~Department of Statistics, Harvard University \vspace{-1mm}}\\
\normalsize{$^\ast$~To whom correspondence should be addressed; email: eckles@mit.edu, rahimian@pitt.edu.}
}

\date{}

\begin{document}

\maketitle

\vspace{-10mm}

\begin{sciabstract}
Network structure can affect when and how widely new ideas, products, and behaviors are adopted. In widely-used models of biological contagion, interventions that randomly rewire edges (on average making them ``longer'') accelerate spread. However, there are other models relevant to social contagion, such as those motivated by myopic best-response in games with strategic complements, in which an individual's behavior is described by a threshold number ($\theta$) of adopting neighbors above which adoption occurs (i.e., complex contagions). Recent work has argued that highly clustered, rather than random, networks facilitate spread of these complex contagions. Here we show that minor modifications to this model reverse this result, thereby harmonizing qualitative facts about how network structure affects contagion. To model the trade-off between long and short ties, we analyze the rate of spread over networks that are the union of circular lattices and random graphs on $n$ nodes. Allowing for noise in adoption decisions (i.e., adoptions below threshold) to occur with order ${n}^{-1/\theta}$ probability along at least some ``short" cycle edges is enough to ensure that random rewiring accelerates the spread of a noisy threshold-$\theta$ contagion. This conclusion also holds under partial but frequent enough rewiring and when adoption decisions are reversible but infrequently so, as well as in high-dimensional lattice structures that facilitate faster-expanding contagions. Simulations illustrate the robustness of these results to several variations on this noisy best-response behavior. Hypothetical interventions that randomly rewire existing edges or add random edges (versus adding ``short'', triad-closing edges) in hundreds of empirical social networks reduce time to spread. This revised conclusion suggests that those wanting to increase spread should induce formation of long ties, rather than triad-closing ties. More generally, this highlights the importance of noise in game-theoretic analyses of behavior.
\end{sciabstract}

\begin{figure}[t]
\begin{center}
\hspace{-15pt}
\begin{subfigure}[b]{0.5\columnwidth}
\stackinset{c}{.33\textwidth}{t}{.3\textwidth}
   {\includegraphics[width=0.2\linewidth]{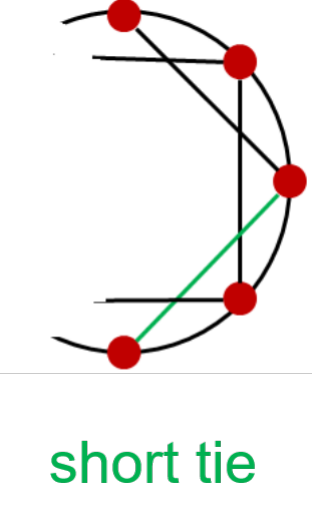}}
   {\includegraphics[width=1.1\linewidth]{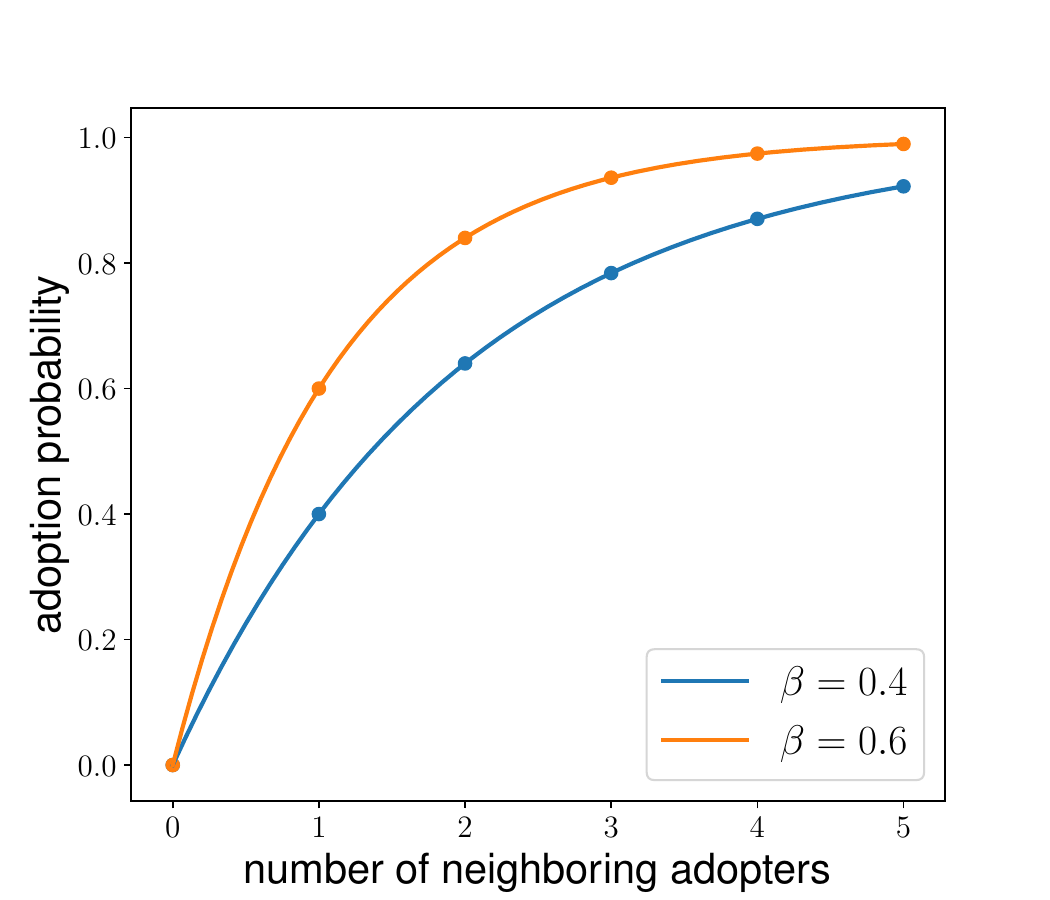}}
   \caption{Simple Activation Functions}
\label{fig:simple_activation_functions} 
\end{subfigure}~\hspace{-5pt}
\begin{subfigure}[b]{0.5\columnwidth}
\stackinset{c}{-.3\textwidth}{t}{.325\textwidth}
   {\includegraphics[width=0.18\linewidth]{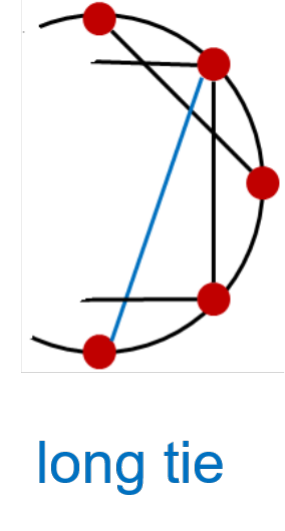}}
   {\includegraphics[width=1.1\linewidth]{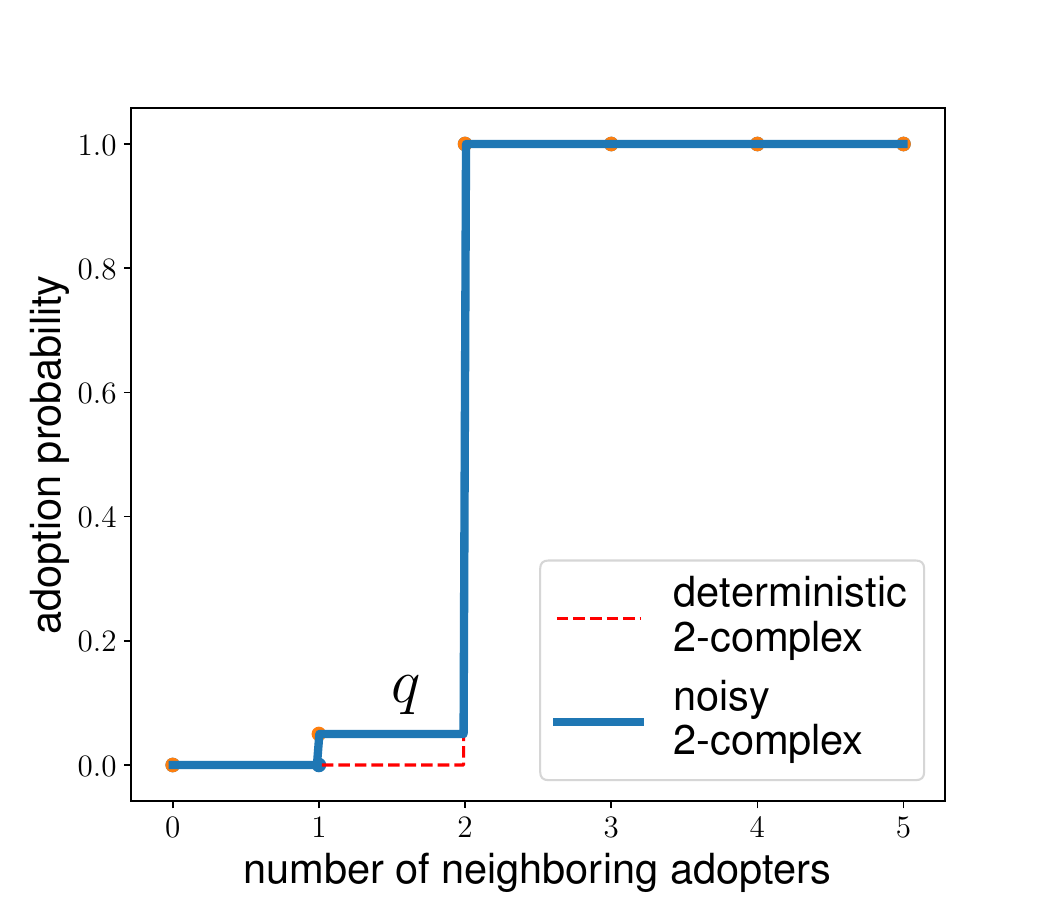}}
   \vspace{-15pt}

\caption{Complex Activation Functions}
\label{fig:complex_activation_functions}
\end{subfigure}
\end{center}
\vspace{-10pt}
\caption{\small{Activation functions for (A) simple contagion and (B) variations on complex contagion. In the case of a simple activation function (A), every edge has an independent probability $\beta$ of transmitting infections (adoptions); subsequently, the probability of adoption with $x$ adopters in the social neighborhood is given by $1 - (1-\beta)^x$. In the case of a noisy threshold-based contagion model (B), there is a non-zero probability ($q>0$) of adoptions below threshold. The inset figures illustrate rewiring (A) a short tie to get (B) a long tie, on a $4$-regular circular lattice structure which we refer to as cycle-power-$2$ and denote by  $\mathcal{C}_2$. Rewiring  a short tie speeds up the spread of simple contagions: \emph{strength of the weak (or long) ties}. Recent studies have arrived at an opposite conclusion for complex contagion: \emph{weakness of long ties}.}}
\vspace{-10pt}
\label{fig:activation_functions}
\end{figure}

\newpage
\noindent
How does network structure affect the spread of ideas, products, and behaviors? Social interactions among individuals facilitate a diverse range of contagions, and understanding the role of contact structure is central to the social and behavioral sciences. Decision-makers often rely on their knowledge of contagion in planning interventions that seed a behavior \cite{leskovec2007dynamics, kempe2003maximizing, hinz2011seeding, libai2013decomposing, beaman2018can}, prevent or reverse infection of nodes \cite{cohen2003efficient, preciado2014optimal, chami2017social}, or that attempt to modify network structure \cite{chaoji2012recommendations, valente2012network, carrell2013natural, cerdeiro2017individual}. Unfortunately, existing analyses of the two most widely used families of models (\emph{simple and complex contagions}) have led to opposing conclusions (\emph{weakness or strength of long ties}) about how network structure --- in particular, clustering --- affects spread of behavior; see Figure~\ref{fig:activation_functions} insets.

Social contagions that are expected to be driven by incidental transfer of information are often modeled analogously to biological contagion of infectious disease. In such \emph{simple contagion} models, a node has an independent (and typically identical) probability of being infected by each infected neighbor \cite{dodds2005generalized}; see Figure~\ref{fig:activation_functions}A. It is well known that such contagions spread more slowly in highly clustered networks than in more random networks \cite{watts1998collective,hebert2010propagation}.
Related considerations lead to the ``strength of weak ties'' hypothesis by which ``weak'' (or more properly ``long'') ties play critical roles in access to valuable information \cite{granovetter1973strength,aral2011diversity,jahani2023long}, such as in labor markets \cite{gee2017paradox, rajkumar2022causal}. On the other hand, adoptions which are costly, or occur because of normative social pressure or coordination, are often modeled as myopic best-responses in repeated graphical games of strategic complementarities (such as coordination games), whereby nodes' utilities from adopting depend on the number of adopting neighbors \cite{galeotti2010network,blume1993statistical, morris2000contagion,young2011dynamics}. Threshold activation functions are the archetypal example of such \emph{complex contagion} models \cite{granovetter1978threshold,centola2007complex}. In their canonical form, a single parameter $\theta$ divides non-adoption from adoption such that adoption occurs if and only if the number of adopting neighbors reaches the threshold $\theta$ (Figure \ref{fig:activation_functions}B); call this \emph{deterministic $\theta$-complex contagion}.
Recent analyses of this deterministic model \cite{centola2007complex}, or limits of noisy best responses as the noise level goes to zero ($q \to 0$ in Figure \ref{fig:activation_functions}B) \cite{montanari2010spread}, have concluded that the spread of complex contagions is facilitated by more clustered networks, emphasizing the role of short ties in formation of wide bridges or complex paths \cite{guilbeault2021topological}, such that there is a ``weakness of long ties'' \cite{centola2007complex}. These lead to opposite recommendations about how to intervene on a network to facilitate (or slow) a social contagion.

Is such a deterministic model robust to modeling variations that may make it more consistent with both empirical evidence and widely-used random utility models of choice? Empirical studies of social contagion, including those that provide evidence for complex contagion, find substantial probability of adoption with a single adopting neighbor \cite{leskovec2007dynamics,bakshy2012role,bakshy2012social,centola2010spread}, and empirical adoption rates with an additional adopter (beyond the first) increase by less than a factor of five (SI Figure S2), suggesting substantial nondeterminism or heterogeneity. More generally, rather than positing determinism, analyses of discrete choice problems typically hypothesize that individuals are random utility maximizers, thereby leading to positive choice probabilities specified by, e.g., probit or logit functions over the entire support~(SI section S4.3). Here we show that allowing a small probability of below-threshold adoption (denoted by $q$ in Figure~\ref{fig:activation_functions}B), even only via some short ties, reverses existing stylized facts about how network structure affects the spread of complex contagions, putting the emphasis back on the important structural role of long ties for complex contagions (see SI Figure S5). This harmonizes theoretical guidance about how network structure affects the spread of both simple and complex contagions.

\section*{Results}

\begin{figure}[tb]
\begin{center}
\includegraphics[width=0.25\linewidth]{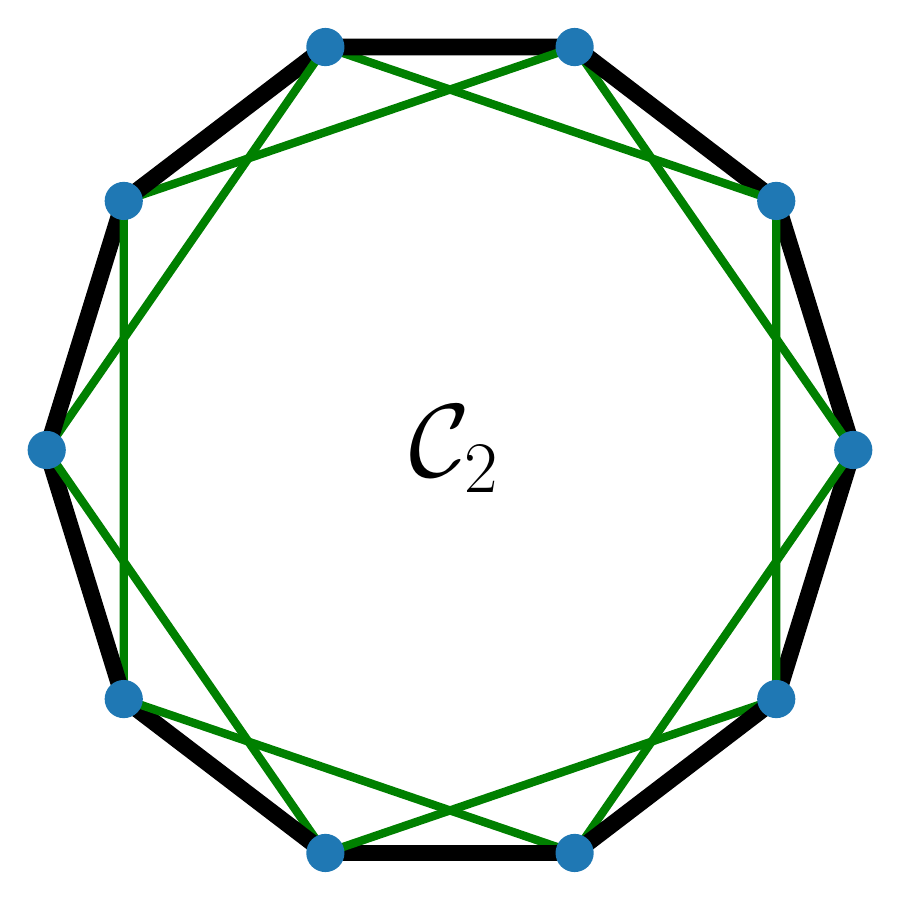}\includegraphics[width=0.25\linewidth]{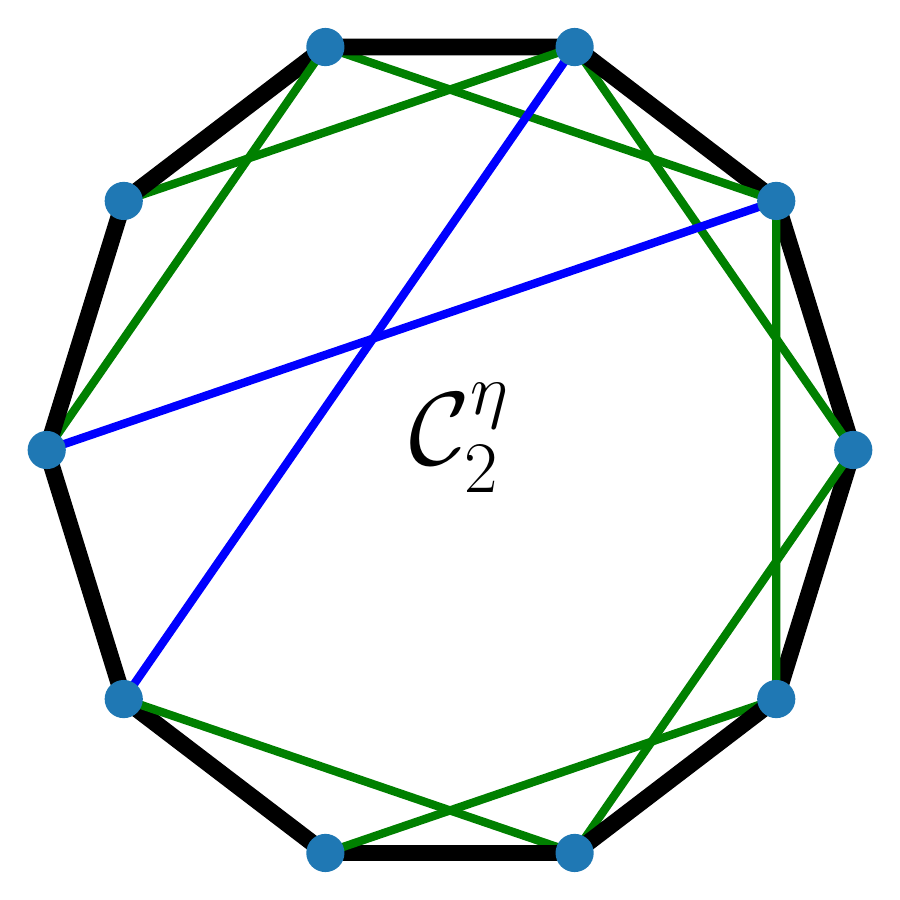}\includegraphics[width=0.25\linewidth]{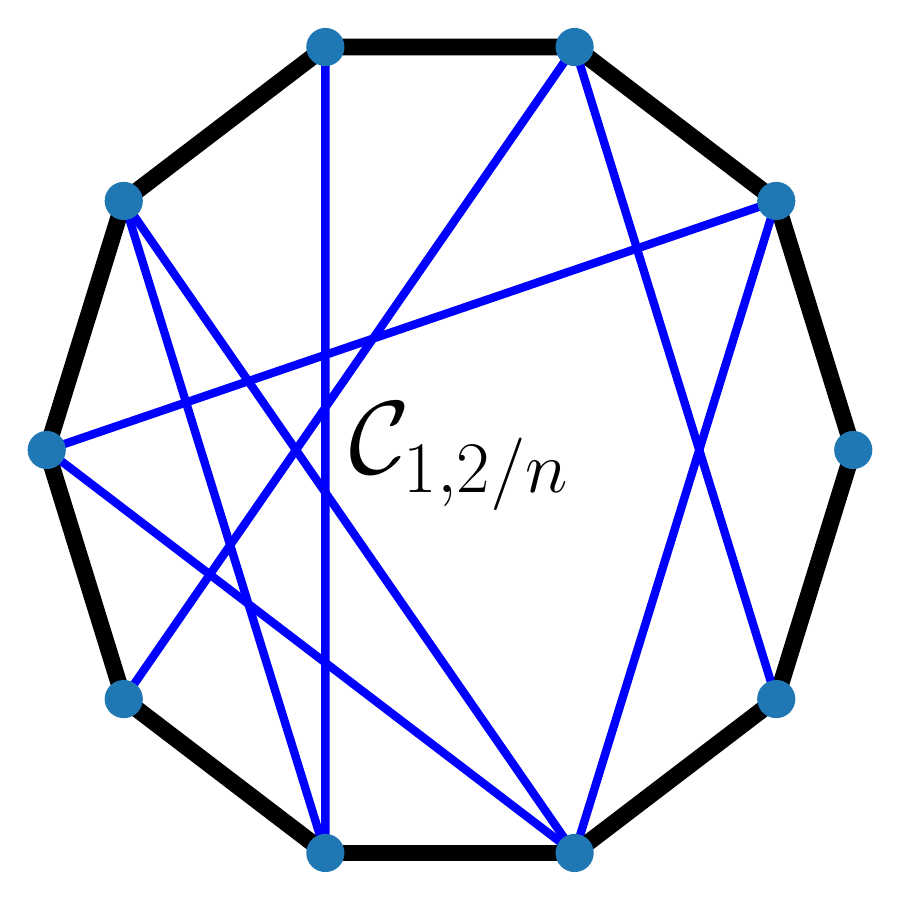}
\end{center}
\caption{\small{We consider the rewiring of a cycle-power-$2$ graph ($\mathcal{C}_2$, left) as $\mathcal{C}_2\setminus\mathcal{C}_1$ edges (short ties in green) are removed and replaced by random edges (long ties in blue), keeping the $\mathcal{C}_1$ edges (in black) fixed. Deterministic $2$-complex contagion takes exactly $\lfloor n/2 \rfloor$ steps (i.e., $n/2-1$ for $n$ even and $(n-1)/2$ for $n$ odd) to spread over the entire $\mathcal{C}_2$ graph because starting from a pair of neighboring infected nodes at each step two new nodes are going to be infected, except for the last step in odd-sized networks where the contagion ends with the conversion of the single remaining node. Noisy $2$-complex contagion spreads faster due to the additional possibility of passing infections through single, infected nodes, however, even if all nodes with infected neighbors become infected at every time step, it still takes $\lfloor n/4 \rfloor$ steps for contagion to spread across $\mathcal{C}_2$ entirely. On the other hand, a deterministic $2$-complex contagion does not spread totally on the cycle-union-random-graph ($\mathcal{C}_{1,2/n}$ in the middle). In Theorem \ref{thm:C_1unionGn}, we bound the spread time of noisy $2$-complex contagion on $\mathcal{C}_{1,2/n}$ and show that when probability of adoptions below threshold is large enough, the noisy complex contagion spreads faster in $\mathcal{C}_{1,2/n}$ compared to $\mathcal{C}_2$. In Theorem \ref{thm:interpolatingC_1}, we interpolate between $\mathcal{C}_2$ and $\mathcal{C}_{1,2/n}$ by rewiring the edges on $\mathcal{C}_2\backslash\mathcal{C}_1$, keeping the average degree of the nodes constant (equal to four). Using $\eta$ to denote the expected number of rewired edges ($\mathcal{C}_{2}^{\eta}$ on the right), we show that for $q$ and $\eta$ large enough noisy $2$-complex contagion spreads faster on $\mathcal{C}_{2}^{\eta}$ than $\mathcal{C}_{2}$. In Theorem \ref{thm:noninertial}, we study a variation of the noisy $2$-complex contagion by allowing infected nodes to revert to the susceptible state with a probability $\delta$. For this reversible, noisy, $2$-complex contagion we again show a faster spread on $\mathcal{C}_{1,2/n}$ than $\mathcal{C}_2$, for $q$ large and $\delta$ small enough.}}
\vspace{-10pt}
\label{fig:interpolation_theorems_and_mechanism_for_spread}
\end{figure}

For our analytical results, we study the spread of $\theta$-complex contagion starting from $\theta$ adjacent infected nodes in a variation on ``small world'' networks \cite{watts1998collective}. More specifically, we begin with $2$-complex contagion from a pair of adjacent adopter nodes on rewired, circular, lattice networks. We use $\mathcal{C}_{k}$ to denote a cycle-power-$k$ graph which is defined as a circular lattice on $n$ nodes where each node is connected to its $2k$ nearest neighbors on the cycle (Figure \ref{fig:interpolation_theorems_and_mechanism_for_spread}); the case $k=1$ corresponds to an ordinary cycle ($\mathcal{C}_{1}$). In preliminary analysis (SI section S3), we show that the spread time of deterministic $2$-complex contagion on $\mathcal{C}_2$-union-random-graph is with high probability upper-bounded by $2 n^{2/3} (\log \log n)^2$, which is asymptotically faster than the spread time on any cycle power graph (the latter being of order $n$). Hence, rewiring short ties and replacing them with random, long ties speed up the spread of $2$-complex contagion on a cycle power graph even without below-threshold adoptions (see SI Figures S3--S4). That preliminary result characterizes how long ties can speed up complex contagions, but only when the required short-tie structure for deterministic, $2$-complex contagion (i.e., $\mathcal{C}_2$), is intact. Here we focus on what happens when $\mathcal{C}_2$ edges are replaced with random edges (i.e., \emph{long ties}) so that a deterministic, $2$-complex contagion would not spread on the rewired graph. Instead, we consider a \emph{noisy} $2$-complex contagion that allows for a non-zero (but vanishing as $n\to\infty$) probability of \emph{simple} adoptions. We denote this probability of adoptions below threshold by $q$ --- or $q_n$ to emphasize it dependence on the network size in analytical results. Formally, this implies that nodes that have only one infected neighbor get infected independently in each round with probability $q_n$.

Let $\mathcal{C}_{1,c/n} := \mathcal{C}_1 \cup \mathcal{G}_{n, c/n}$, where $\mathcal{G}_{n,c/n}$ is an Erd\H{o}s--R\'{e}nyi random graph with edge probability $c/n$ for some fixed constant $c>0$. In case of $\mathcal{C}_{1,2/n}$, the union graph has $n$ nodes, each node has expected degree four ($4 -2/n$,  to be precise), and the set of edges is the union of edges in the two graphs. Let ${T}_{\theta,q}(\mathcal{G})$ be the random variable representing the total spread time of noisy $\theta$-complex contagion with simple adoption probability $q$ over graph instance $\mathcal{G}$. The following theorem upper-bounds  ${T}_{2,q}(\mathcal{C}_{1,2/n})$ by $(4\sqrt{n}/q_n) (\log \log n)^2$ implying that if $\sqrt{n}q_n \to \infty$, then noisy $2$-complex contagion spreads faster on $\mathcal{C}_{1,2/n}$ than $\mathcal{C}_2$: ${T}_{2,q}(\mathcal{C}_{1,2/n}) \ll {T}_{2,q}(\mathcal{C}_{2})$ for $1/\sqrt{n} \ll q$. Throughout, we use $\ll$ to indicate asymptotic dominance: $a_n \ll b_n$ iff $a_n/ b_n \to 0$ as $n\to\infty$, and all our results hold with high probability (w.h.p.), i.e., with probability tending to one as $n \to \infty$. Note that (see Figure \ref{fig:interpolation_theorems_and_mechanism_for_spread} caption): $\lfloor n/4 \rfloor \leq {T}_{2,q}(\mathcal{C}_{2}) \leq \lfloor n/2 \rfloor$.

\begin{Th}
\label{thm:C_1unionGn} 
Consider the noisy $2$-complex contagion over $\mathcal{C}_{1,2/n}$ with simple adoption probability $q$. With high probability as $n \to \infty$, ${T}_{2,q}(\mathcal{C}_{1,2/n}) < \frac{4 \sqrt{n}}{q} (\log \log n)^2$.
\end{Th}

The proof of Theorem \ref{thm:C_1unionGn} provides a clear intuition for how long ties accelerate noisy complex contagions (SI section S4). Recall our $2$-complex contagions are initialized from two infected neighboring nodes on $\mathcal{C}_1$. We upper-bound the spread time by breaking down the contagion into two sub-processes: \textit{(i)} spreading along the cycle ($\mathcal{C}_1$) via rare, sub-threshold adoptions, and \textit{(ii)} above-threshold adoption along the random, long ties. Initially, the infection spreads along the cycle. Once the infected nodes form long enough intervals, the infection spreads to far away points along pairs of random ties. The latter occurs when a susceptible node has at least two long ties connecting it to infected nodes. Our analysis (SI section S4) indicates that an infected interval of length $\sqrt{n}\log\log n$ is long enough to have a pair of random ties be incident to the same node across the cycle (see Figure S5), thus passing the $2$-complex contagion with high probability as $n \to \infty$. This analysis (see SI Theorem 11) shows that the rate of spread over cycle-union-random-graph structures is determined by the time that it takes for the infected intervals along the cycle to grow long enough, to make the spread of complex contagion through their long ties a probable event.


This idea is reasonably general and can be applied to modeling variations where above-threshold adoption occurs with a probability $\rho$ less than one or the adoption thresholds are greater than two ($\theta>2$). In both cases the spread will be slowed down either to wait for above-threshold adoption to occur at the slower $1/\rho$ rate or for the intervals to grow longer to make a more stringent $\theta$-complex adoption through $\theta$ random ties probable. These and other modeling variations are explored with extensive simulations in SI section S9; see Figures S15 to S18. In particular, for noisy $\theta$-complex contagion we have (proved in SI Section S5):
\begin{Th}
\label{Thm:ThetaComplex} 
Consider the noisy $\theta$-complex contagion with simple adoption probability $q$ over $\mathcal{C}_{1,c/n}$ for constant $c\geq\theta$. With high probability as $n \to \infty$, ${T}_{\theta,q}(\mathcal{C}_{1,c/n}) < \frac{4 n^{1-1/\theta}}{q}  (\log \log n)^2$.
\end{Th}


Starting with an infected interval of length $\theta$, it takes $\lceil{(n-\theta)}/{2}\rceil$ time steps for $\theta$-complex contagion to spread on $\mathcal{C}_{\theta}$ entirely. In case of noisy $\theta$-complex contagion, even if all simple contagion adoption attempts at every time step are successful, the total spread will take more than $\lceil{(n-\theta)}/{(2\theta)}\rceil$ steps: ${T}_{\theta,q}(\mathcal{C}_{\theta}) > {(n-\theta)}/{(2\theta)}$. Choosing $q \gg n^{-1/\theta} (\log \log n)^2$ will make ${T}_{\theta,q}(\mathcal{C}_{1,c/n}) \ll {(n-\theta)}/{(2\theta)} < {T}_{\theta,q}(\mathcal{C}_{\theta})$; thence the requisite noise level for achieving a faster spread on the rewired lattice is $n^{-1/\theta} (\log \log n)^2$ which increases with increasing $\theta$. With heterogeneous thresholds, let us denote the vector of individual thresholds by $\bar{\theta} = (\theta_1,\ldots, \theta_n)$ with ${\theta}_{\max} = \max\{\theta_i, i \in [n]\}$, and let ${T}_{\bar{\theta},q}(\mathcal{G})$ be the random variable measuring the total spread time of noisy ``$\bar{\theta}$-complex contagion" with heterogeneous threshold vector $\bar{\theta}$ and simple adoption probability $q$ over a (random) graph instance $\mathcal{G}$. Note that decreasing each individual's threshold from ${\theta}_{\max}$ to $\theta_i$ for $i \in [n]$ can only speed up the spread, hence, ${T}_{\bar{\theta},q}(\mathcal{C}_{1,c/n}) \leq {T}_{\theta_{\max},q}(\mathcal{C}_{1,c/n})$ for any realization of random graphs and simple contagion infections (i.e., point-wise over the probability space). Subsequently, Theorem~\ref{Thm:ThetaComplex} is directly applicable to the case of heterogeneous thresholds after replacing $\theta$ with $\theta_{\max}$: setting the simple adoption probability $q \gg n^{-1/\theta_{\max}} (\log \log n)^2$ is sufficient to ensure a faster noisy $\bar{\theta}$-complex contagion spread over the rewired lattice because then ${T}_{\bar{\theta},q}(\mathcal{C}_{1,c/n}) \leq {T}_{\theta_{\max},q}(\mathcal{C}_{1,c/n}) \ll n$, with high probability as $n\to\infty$. 

In what follows we investigate the robustness of our analytical findings in three respects: \emph{(i)} when there are only some edges rewired (Theorem \ref{thm:interpolatingC_1}), \emph{(ii)} when the adoptions are reversible (Theorem \ref{thm:noninertial}), and \emph{(iii)} when contagions spread in high-dimensional structures (SI Section  S8). Further robustness is revealed by simulations with other modeling variations and with empirical networks (SI Section S9).

\vspace{10pt}
\noindent \textbf{Rewiring only some edges.}
We begin by interpolating continuously between $\mathcal{C}_2$ and the random graph $\mathcal{C}_{1,{2}/{n}}$. We track the evolution of the spreading time along the interpolation path, using a random graph model $\mathcal{C}^{\eta}_2$ and study what happens as the edges in $\mathcal{C}_2$ but not in $\mathcal{C}_1$ (i.e.~$\mathcal{C}_2\setminus\mathcal{C}_1$) are rewired. In this context, $\eta$ denotes the ``expected'' number of edges that are rewired to construct the random graph $\mathcal{C}_2^{\eta}$ from $\mathcal{C}_2$.  Theorem \ref{thm:interpolatingC_1} upper bounds the total spread time of noisy $2$-complex contagion over $\mathcal{C}_2^{\eta}$ for $\eta>\sqrt{n}$.

\begin{Th}\label{thm:interpolatingC_1} Consider a noisy $2$-complex contagion over $\mathcal{C}^{\eta}_2$ with simple adoption probability $q$. Let $\eta = n^{\nu}$ and $\frac{1}{2}<\nu<1$, then with high probability as $n\to\infty$, ${T}_{2,q}(\mathcal{C}_2^{\eta}) < 4(\sqrt{n}/q + n^{3/2- \nu})(\log \log n)^2$. 
\end{Th}

\begin{figure}[H]
\begin{center}
\begin{subfigure}[b]{0.5\textwidth}
   \includegraphics[width=1\linewidth]{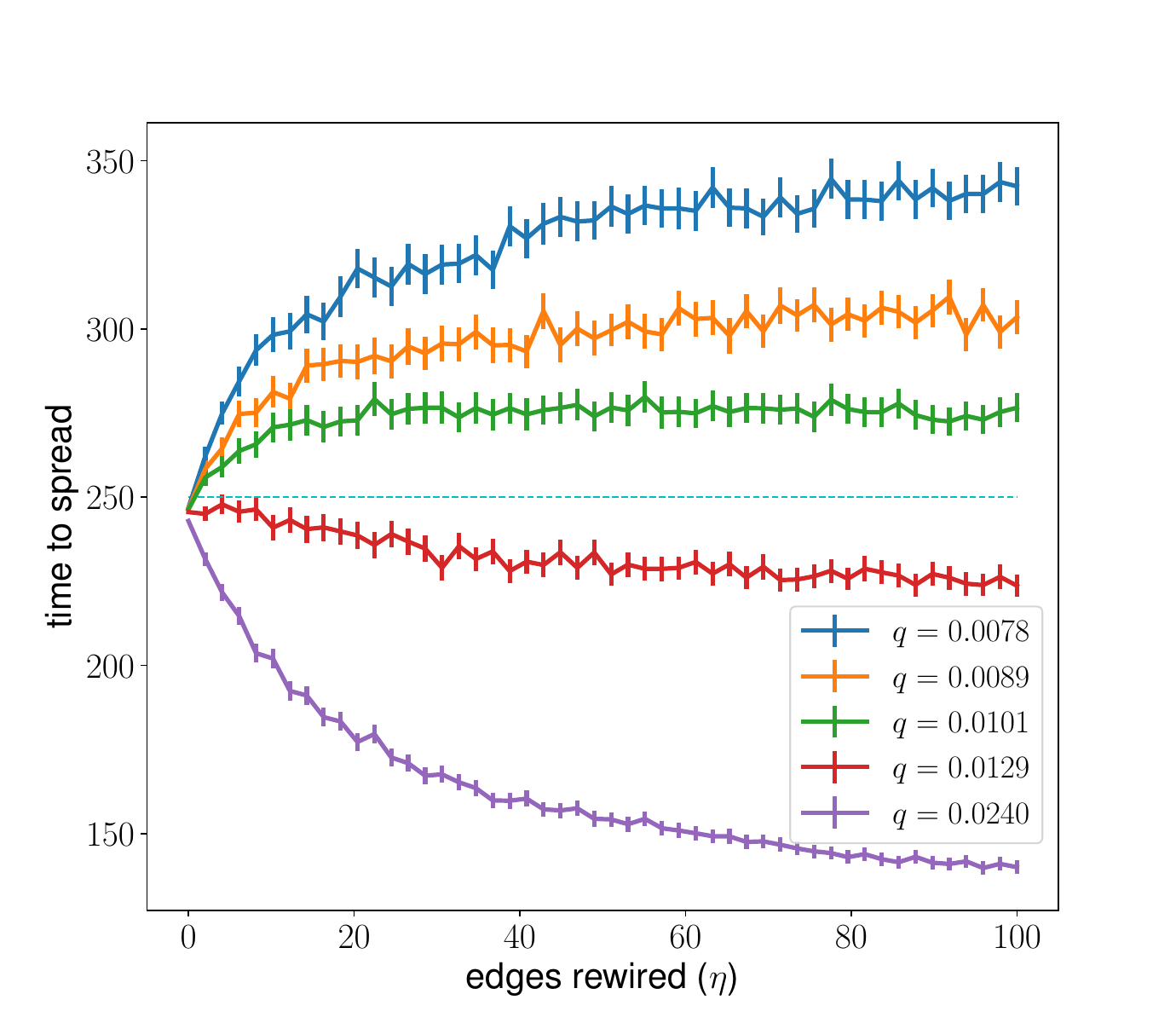}
   \caption{$T_{2,q}(\mathcal{C}_2^{\eta})$ versus $\eta$}
   \label{fig:eta}
\end{subfigure}\hspace{-15pt}~\begin{subfigure}[b]{0.5\textwidth}
   \includegraphics[width=1\linewidth]{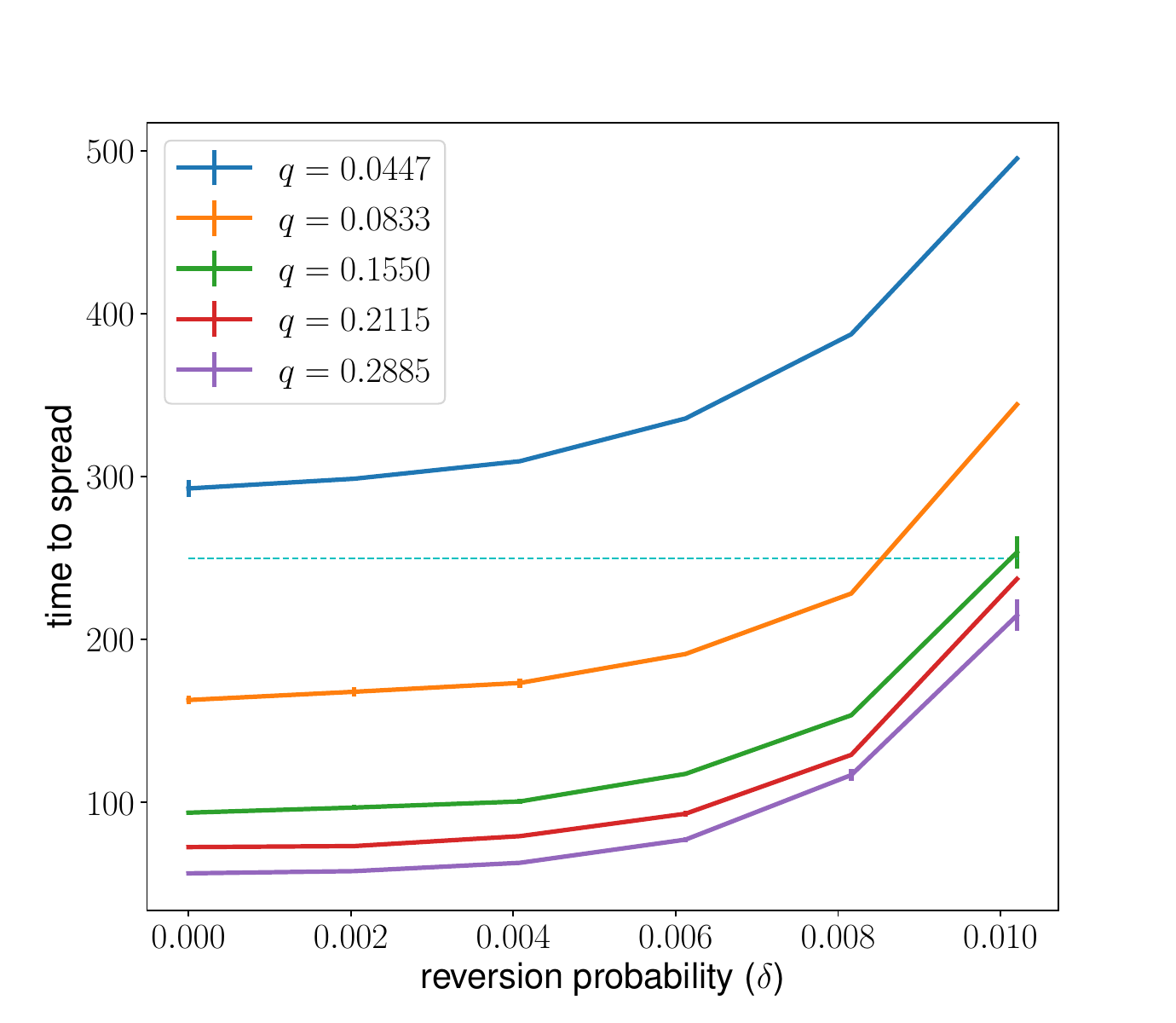}
   \caption{$T^{\delta}_{2,q}(\mathcal{C}_{1,2/n})$ versus $\delta$}
   \label{fig:delta}
\end{subfigure}
\end{center}
\vspace{-20pt}
\caption{\small{Spreading time of noisy complex contagion over rewired $\mathcal{C}_2$ graphs. In (\subref{fig:eta}), we follow the same model as in Theorem \ref{thm:interpolatingC_1}: noisy $2$-complex contagion with sub-threshold adoption probability $q$. In (\subref{fig:delta}), we follow the reversible noisy $2$-complex contagion model of Theorem \ref{thm:noninertial} with reversion probability $\delta$  and sub-threshold adoption probability $q$. All networks have $n = 500$ nodes. The the spread time of $2$-complex contagion on $\mathcal{C}_2$ is $250$ which is marked by a dashed line. Each point is the average of $1000$ random draws. The vertical bars indicate the $95\%$ normal confidence intervals around the means.}}
\label{fig:fixed_degree_unions}
\end{figure}

In Theorem \ref{thm:interpolatingC_1}, we parameterize $\eta = n^{\nu}$ and for $\nu \in (\frac{1}{2},1)$ we upper-bound the spread time of noisy $2$-complex contagion by $n^{3/2 - \nu} + \sqrt{n}/q$ (disregarding the logarithmic factors). For $\nu$ large enough ($\nu\to1$) and $q \ll 1$, $\sqrt{n}/q$ is the dominant term that fixes the spread time independently of $\nu$, and we recover the $({4 \sqrt{n}}/{q}) (\log \log n)^2$ upper bound in Theorem \ref{thm:C_1unionGn}. However, for $1/\sqrt{n} \ll q$ we can specify a range of $\nu$ for which increasing $\nu$ decreases the upper bound, with complex contagion spreading through the long ties. In particular, if  $q = n^{-1/2+\nu'}$, then for $\frac{1}{2}<\nu<\frac{1}{2}+ \nu'$  contagion spreads faster in $\mathcal{C}^{\eta}_2$ compared to $\mathcal{C}_2$. In comparison, the spread time on $\mathcal{C}_2$ is at least $\lfloor n/4 \rfloor$; see Figure \ref{fig:interpolation_theorems_and_mechanism_for_spread} caption. Figure \ref{fig:eta} shows the spreading time versus the rewiring parameters $\eta$ for $\mathcal{C}_2^{\eta}$ random graphs. It confirms that when the probability of adoptions below threshold ($q$) is large enough, the more we rewire $\mathcal{C}_2\setminus\mathcal{C}_1$ edges, the faster the noisy $2$-complex contagion. On the other hand, when $q$ is very small, rewiring slows down the spread as it blocks the passage of fast $2$-complex contagion along $\mathcal{C}_2$. Our theoretical analysis in SI Section S6.1 shows that in the $\eta \ll \sqrt{n}$ regime, there are not enough long ties to initiate $2$-complex contagion across the cycle and increased rewiring slows down the spread (see Theorem 24 in SI Section S6.1).

\vspace{10pt}
\noindent\textbf{Reversible adoption.}
While prior work has examined ``stochastic thresholds'' via simulations \cite[p. 724]{centola2007complex}, that model importantly differs from our noisy threshold-based contagions as the activation function in that work (although allowing for sub-threshold adoptions) is reevaluated at every time step without regard for each nodes' own prior adoption. That is, an infected node readily reverts back to being susceptible after reevaluation of its activation function, leading to a perfectly reversible, non-inertial process. This significantly diminishes the effect of sub-threshold adoptions because nodes that are activated below threshold are likely to be stochastically turned off when their activation function is reevaluated.

Indeed, allowing sub-threshold adoptions in a perfectly reversible complex contagion is not enough to reverse the conclusions about the weakness of long ties \cite{centola2007complex}: some inertia is critical to our conclusions. We note that many of the decisions theorized to be governed by threshold-based contagions have some inertia, often because they are costly or could be difficult to reverse (e.g., purchases). In the mechanism that we have identified for the spread of noisy $2$-complex contagion on $\mathcal{C}_{1,{2}/{n}}$, it is critical to allow simple contagion to spread along an interval on $\mathcal{C}_{1}$, until two random (long) ties are likely to be adjacent to the infected interval (Figure S5). Reversions at the interval boundaries can disrupt this interval growth mechanism and significantly slow down the spread of complex contagion through long ties. Notwithstanding, we can allow infected nodes to revert back to being ``susceptible" with probability $\delta$, and for small $\delta$ our conclusions hold. Note that in this model even with all nodes infected, a fraction $\delta$ of them are likely to revert to being susceptible. Therefore, it is useful to define $T_{2,q}^{\delta}(\mathcal{C}_{1,2/n})$ as the first time that $n(1-\delta)$ nodes are infected starting from two neighboring infected nodes on the cycle under a reversible, noisy $2$-complex contagion model with sub-threshold adoption probability $q$ and reversion probability $\delta$. Our following result provides an upper bound on $T_{2,q}^{\delta}(\mathcal{C}_{1,2/n})$, extending Theorem \ref{thm:C_1unionGn} to allow for a positive probability of reversions $\delta>0$.

\begin{Th}\label{thm:noninertial} Consider a reversible, noisy $2$-complex contagion over $\mathcal{C}_{1,2/n}$ with simple adoption probability $q$ and reversion probability $\delta$. If $\delta  \ll 1/\sqrt{n}$, then  $T_{2,q}^{\delta}(\mathcal{C}_{1,2/n}) < \frac{4 \sqrt{n}}{q-\delta} (\log \log n)^2$, with high probability as $n \to\infty$.
\end{Th}
Of note, the spread time measured by $T_{2,q}^{\delta}(\mathcal{C}_{1,2/n})$ upper-bounds the faster model where each infected agent reevaluates its activation function with probability $\delta$, rather than reverts to being susceptible. In the former case, we recover the stochastic threshold model of prior work \cite{centola2007complex} as $\delta \to 1$. Similarly, we can recover Theorem \ref{thm:C_1unionGn} by letting $\delta\to0$. Figure \ref{fig:delta} illustrates the effect of $\delta$ on the spread time with numerical simulations.

\vspace{10pt}
\noindent\textbf{Higher-dimensional lattices.}
Contagion on circular lattices is limited to a single dimension. In higher dimensions, spread speeds up significantly as contagion expands simultaneously across multiple dimensions. Our analysis of noisy complex contagions on circular lattice such as $\mathcal{C}_2$ and $C_{1,2/n}$ has natural extensions in $d$ dimensions. In SI Section S8, we formalize this for $d=2$ and show that contagion on a $\sqrt{n}\times\sqrt{n}$ square lattice with random, long-range ties can be upper bounded by $({36 n^{1/4}}/{q})  (\log \log n)^{3/2}$, which for $q \gg 1/n^{1/4}$ is strictly faster than the spread time on the square lattice with closed triads, the latter being  of the order $\sqrt{n}$. Generally, for $q\gg 1/n^{1/2d}$ the spread time on $d$-dimensional hypercube with random, long-range ties is strictly faster than ${n}^{1/d}$ which is the order of time that noisy $2$-complex contagion takes to  spread on the $d$-dimensional hypercube with closed diagonals (i.e., triad-closing ties).

\subsection*{Empirical networks}
Our theoretical analysis of lattice structures with random, long-range ties is motivated by the well-documented, small-world phenomena in real social networks that simultaneously exhibit the high clustering of regular latices and short average path length of random graphs \cite{watts1998collective}. The presence of long ties in real social networks could be due to life events \cite{jahani2023long} or through strategic efforts of network participants \cite{jackson2005economics}, and such ties are generally associated with positive economic outcomes. Social network platforms can also actively shape the network structure through link recommendations, e.g., LinkedIn’s ``People You May Know'' algorithm, with measurable consequences for individuals and groups \cite{rajkumar2022causal}. In our simulation studies of contagions on empirical social networks, we test the effect of not only rewiring existing links but also adding either new random or new triad-closing edges to better inform interventions by platforms and others.

Our simulation results on empirical networks and with a broad class of contagion models support the robustness of our theoretical findings. We use five sets of empirical social networks; see SI section S9, for a description of each. For each social network, contagion begins from two adjacent random seeds, and we measure the time to $90\%$ spread under four conditions: (i) the original networks (no intervention), (ii) with $10\%$ of edges rewired, (iii) with $10\%$ added edges selected proportional to the number of triads they close, and (iv) with $10\%$ new edges added randomly. For each intervention type, we simulate the spread times over the modified networks $500$ times. In these simulations, a node adopts with certainty if it has at least two adopter neighbors. We fix the probability of adoption with a single adopter neighbor at $q = 0.05$.

\begin{figure}[H]
\centering
\vspace{-1cm}
\begin{subfigure}[b]{.49\columnwidth}
   \includegraphics[width=\columnwidth]{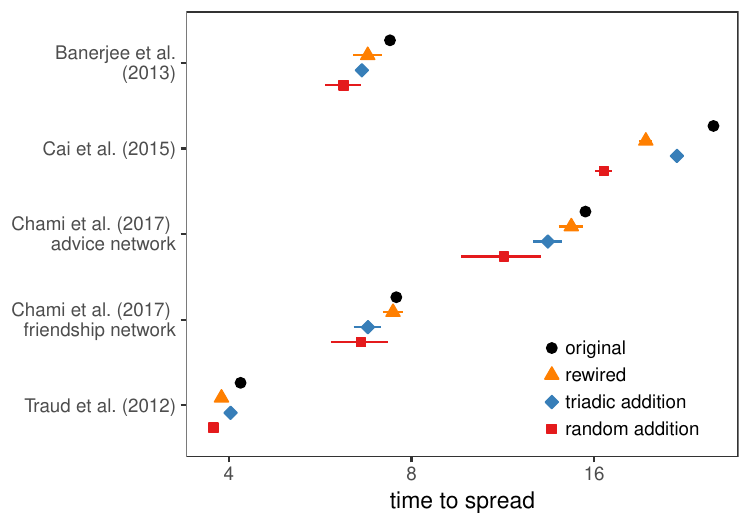}
   \caption{Mean spread times}
   \label{fig:mean_spread_times}
\end{subfigure}~ 
\begin{subfigure}[b]{.49\columnwidth}
   \includegraphics[width=\columnwidth]{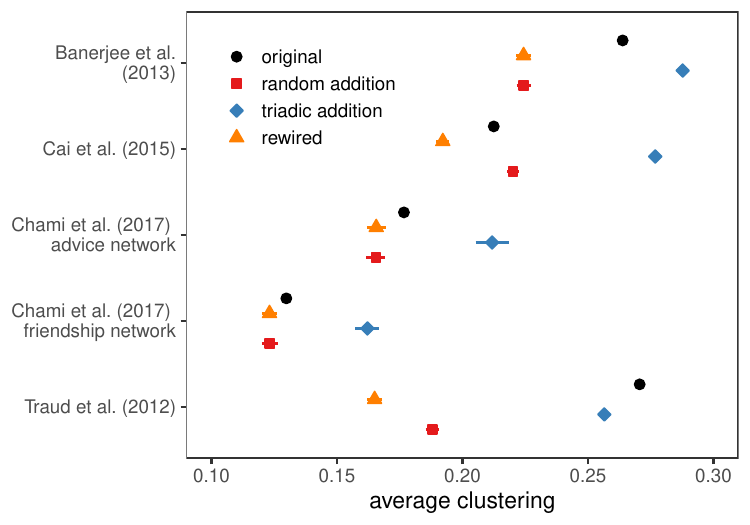}
   \caption{Mean network average clustering}
   \label{fig:avg-clustering}
\end{subfigure} 
\begin{subfigure}[b]{.5\columnwidth}
\includegraphics[width=\columnwidth]{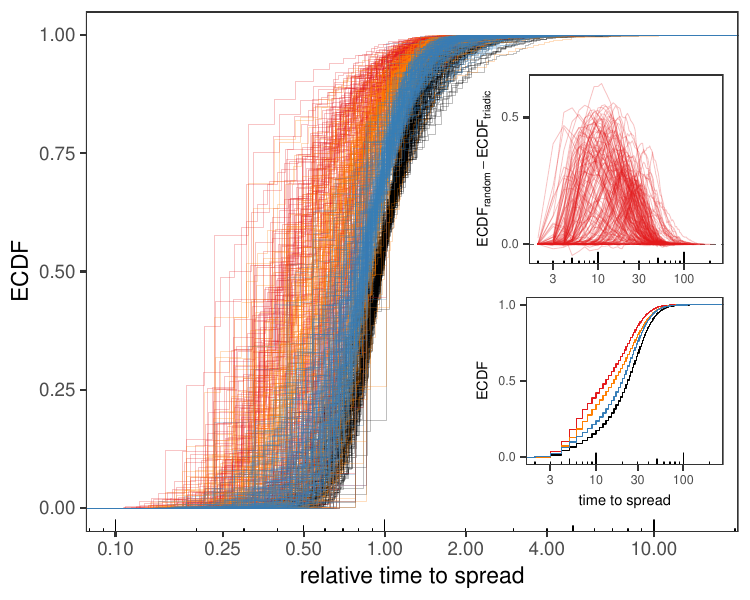}
   \caption{Distributions of spread times}
   \label{fig:ecdf_spread_times}
\end{subfigure}

\caption{\small{(\subref{fig:mean_spread_times}) Mean time to spread in each set of empirical networks. Each point averages over all networks in that set. Error bars are 95\% confidence intervals for the difference from the original network computed by treating each network as a single observation. (\subref{fig:avg-clustering}) The effect that each intervention type has on the network average clustering coefficients. The confidence intervals indicate the the difference from the original network computed by treating each network as a single observation. (\subref{fig:ecdf_spread_times}) Distribution of time to spread for each of the 175 networks of Chinese households in Cai et al. \cite{cai2015social}. For each village, we plot the empirical cumulative distribution function (ECDF) of the spreading times in the original village (black) and under rewiring (orange), as well as random (red) and triad-closing (blue) edge additions. Hence, the main figure overlays $4 \times 175 = 700$ curves, corresponding to the ECDFs of the $500$ spreading time samples computed for each village under the four conditions. Time to spread is normalized by the mean time to spread in the original network. Compared with closing triads, adding random edges consistently speeds up spread, as illustrated by the positive difference in ECDFs (upper inset). The positive difference in each case implies stochastic dominance: the spreading time over the village network with $10\%$ added triad-closing edges dominates (is slower than) the spreading time over the network with the $10\%$ new edges added randomly. The distributions of time to spread averaging over all 175 networks (lower inset) illustrate that both rewiring and random additions speed up the contagion.
}}
\label{fig:empirical_nets}
\end{figure}

Across all five network datasets, random rewiring decreases mean time to spread (Figure \ref{fig:mean_spread_times}). Furthermore, adding random, rather than triad-closing, edges likewise reduces mean time to spread.  To see the aggregate effect that the interventions have on the network structure, we plot the mean values of the network average clustering coefficients under the three interventions compared to the original networks in each dataset (Figure \ref{fig:avg-clustering}). Random addition and rewiring generally decrease the average clustering compared to the original networks. One would expect that triad-closing edge additions should lead to the highest average clustering among the four conditions;however, addition of triad-closing edges can also introduce new open triangles in the vicinity of the added edge. In the case of the Traud et al. (2012) dataset we see a decrease in average clustering with triadic addition. This can point to the existence many broker nodes who connect otherwise disconnected regions of the network. Triadic additions in the neighborhood of the broker nodes will close some triangles while generating many open ones between the newly connected network regions. We further examine spreading times for each network in the largest set --- households in 175 villages in rural China \cite{cai2015social} --- in Figure \ref{fig:ecdf_spread_times}, where we observe a corresponding shift in the distribution of spreading times.

These results with empirical networks are robust to a number of variations, including the intervention size (SI Figure S18) and the percentage of total spread (SI Figure S14). In SI section S9, we present a variation of this model where the probability of adoptions above threshold (called $\rho$) is less than one ($\rho=0.5, q = 0.025$), as well as a case with very small simple adoption probability ($\rho = 1, q = 0.001$). In another variation, infected nodes transition to an inactive state (with probability $\gamma = 0.5$), in which they are no longer infectious, although they are still counted as being adopters. In yet another variation, we consider a fractional threshold model with relative thresholds set to $\theta^{\star}=0.5$. Results are qualitatively similar for other threshold values ($\theta = 2,3,4,5$; SI Figure S15) and heterogeneous thresholds, drawn at random from different distributions (SI Figure S16). Simulation results in all cases reveal the same direction for the effect of interventions, although the effect sizes vary.

\section*{Discussion}
Our simulations indicate that in many real social networks rewiring the edges causes these contagions to spread faster. Moreover, these contagions spread faster when new edges are added uniformly at random rather than with probability proportional to the number of open triads that they close. The latter suggests that it is advantageous to introduce new ties that close fewer triads. This is true even if the decisions to adopt heavily rely on local reinforcement from the neighboring adopters, e.g., with $\rho=1$ and $q=0.001$.

Contrary to the ideas surrounding the ``weakness of long ties''~\cite{centola2007complex, centola2010spread, montanari2010spread}, we find that interventions that introduce long ties via random rewiring or adding random ties accelerate the spread of complex contagions. In common versions of such contagions, there is at least a small probability for adoption to occur even when there is only a single adopter in the social neighborhood. This is enough to change the landscape of results, thereby leading to the conclusion that long ties accelerate these contagions --- just as they do for simple contagions.

Our results indicate that introducing long ties is more effective for accelerating the spread of social contagion --- whether simple and complex. Thus, we propose a more unified recommendation for structural interventions aimed at increasing spread: adding long ties. This conclusion is consistent with empirical studies that identify structural diversity as a correlate of increased adoption \cite{ugander_structural_2012} and document the prevalence of long ties with high information-exchange bandwidth \cite{park2018strength}. 

Interventions in social networks are often unable to directly form arbitrary relationships; rather, they typically consist in some encouragement to interaction. For example, individuals can be randomly assigned to groups, but only some endogenously form friendships, with substantial consequence for the success of such interventions \cite{carrell2013natural}. While there can be noncompliance in edge formation, our simulations suggest that even if one can induce triad-closing edges to form at a greater rate, focusing on forming long ties could still be more effective. In the networks of households in rural China \cite{cai2015social}, we observe that even with $25\%$ additional short, triad-closing ties, spread is slower than with only $10\%$ additional long, random ties (SI Figure S21A). Nonetheless, our results do not address the decision to form a tie. Rather, we clarify the effect that the introduction of new ties has on the speed of spread. We propose the confluence of these two decisions --- whether to form a tie, perhaps in response to an intervention, and whether to adopt a behavior given its adoption by network neighbors --- as a topic for further study.    

A central theme of our work is that deterministic models based on complex contagion are unrealistically constrained and various sources of uncertainty across social networks (observable or unobservable) cause actions in the real world to be significantly stochastic. Capturing the full complexity of network contexts for social contagion, including tie formation tendencies of the agents (homophilous or heterophilous) and timing of their actions, opens up many avenues for future research into the role of network structure. Individuals may show differential preferences for observations that originate from within or outside their communities and structurally-correlated distribution of thresholds can significantly influence which type of social ties are more conducive to contagion. In other contexts individuals may be sensitive to adoptions beyond the local neighborhoods that are coded by pairwise network interactions, leading to a rich variety of contagions across higher-order structures such as hyperedges and simplices~\cite{iacopini2019simplicial,ferraz2023multistability}. In yet other contexts, temporally-nuanced behaviours such as forgetting or burstiness~\cite{akbarpour2018diffusion} may favor special structures, e.g., those that facilitate simultaneous or closely-timed adoptions. We speculate that empirically-grounded theories of social contagion in complex networks can explain rich classes of behaviors and provide a firm foundation for stylized facts about the role of social ties in broad contexts, informing robust interventions that facilitate diffusion of innovations and adoption of new technologies in social networks.


\clearpage 

\section*{Methods} 

\noindent\textbf{Rewiring circular lattices, $\mathcal{C}_{1,2/n}$ and $\mathcal{C}_{2}^{\eta}$.} We use $\mathcal{C}_{1,2/n}$ and $\mathcal{C}_{2}^{\eta}$ to denote graph instances that can be generated by randomly rewiring a cycle-power-$2$ graph on $n$ nodes, keeping the expected degrees of the nodes (asymptotically) fixed at four (Figure \ref{fig:interpolation_theorems_and_mechanism_for_spread}). We construct $\mathcal{C}_{1,2/n}$ by taking the union of an Erd\H{o}s--R\'{e}nyi random graph with edge probability $2/n$ and a cycle $\mathcal{C}_1$. The graph instances of $\mathcal{C}^{\eta}_2$ are meant to continuously interpolate  between $\mathcal{C}_2$, corresponding to $\eta = 0$, and $\mathcal{C}_{1,{2}/{n}}$, corresponding to $\eta \to \infty$. Formally, we define two random graph processes $\mathcal{D}_\eta$ and $\mathcal{G}_{\eta}$ that are coupled through the common index $\eta$. We construct $\mathcal{C}^{\eta}_2$ as a union graph, $\mathcal{C}^{\eta}_2 := \mathcal{C}_1 \cup \mathcal{G}_{\eta} \cup \mathcal{D}_{\eta}$. The coupling between $\mathcal{D}_\eta$ and $\mathcal{G}_{\eta}$ is achieved as follows. To each pair of nodes, $i$ and $j$, we associate independent exponential random variables $X_{ij}$ with mean $n^2$ and $Y_{ij}$ with mean $2n$. Graph $\mathcal{G}_\eta$ is comprised of all edges $\{i,j\}$ for which $X_{ij} < \eta$. Therefore, $\mathcal{G}_{\eta}$ is distributed as Erd\H{o}s--R\'{e}nyi with edge probability $\mathbb{P}\{X_{ij} > \eta\} = 1-e^{-\eta/n^2}$. On the other hand, $D_{\eta}$ is comprised of all edges belonging to $\mathcal{C}_2\setminus\mathcal{C}_1$ for which $Y_{i,j} > \eta$. Therefore, each edge of $\mathcal{C}_2\setminus\mathcal{C}_1$ is missing from $\mathcal{D}_\eta$ with probability $1-e^{-\eta/2n}$, independently of others. In the ${\eta} \ll n$ regime, the expected degree of nodes in $\mathcal{C}^{\eta}_2$ is asymptotically fixed at four, which is the degree of nodes in $\mathcal{C}_2$. Motivated by this observation, we refer to $\mathcal{C}^{\eta}_2$ as the ``$\eta$-rewired $\mathcal{C}_2$'' random graph.\\  

\noindent\textbf{Spreading times ${T}_2$, ${T}_{2,q}$, and ${T}_{2,q}^{\delta}$.} Let $\mathcal{X}_n$ be any graph on $n$ nodes that includes $\mathcal{C}_1$ as a subgraph (e.g., $\mathcal{C}_{1,2/n}$ or $\mathcal{C}_{2}^{\eta}$). We use ${T}_2(\mathcal{X}_n)$, ${T}_{2,q}(\mathcal{X}_n)$, and ${T}_{2,q}^{\delta}(\mathcal{X}_n)$ to denote the random variables measuring the spread times of contagions on $\mathcal{X}_n$ under specific models starting from an infected pair of neighboring nodes on $\mathcal{C}_1$: ${T}_2$ for deterministic $2$-complex contagion, ${T}_{2,q}$ for noisy $2$-complex contagion with simple contagion probability $q$, and ${T}_{2,q}^{\delta}$ for reversible, noisy $2$-complex contagion with simple contagion probability $q$ and reversion probability $\delta$. ${T}_2(\mathcal{X}_n)$ and ${T}_{2,q}(\mathcal{X}_n)$ measure the time until the entire graph $\mathcal{X}_n$ is infected. Deterministic $2$-complex contagion may not spread to the entire $\mathcal{X}_n$, in which case we set ${T}_2(\mathcal{X}_n) = \infty$. ${T}_{2,q}^{\delta}(\mathcal{X}_n)$ measures the first time until $n(1-\delta)$ nodes in $\mathcal{X}_n$ are infected. Our main results in Theorems~\ref{thm:C_1unionGn}-\ref{thm:noninertial} bound these random variables with high probability as $n \to \infty$.\\

\noindent\textbf{Agent-based simulations.} We implement an agent-based model with transitions between susceptible and infected states according to an activation function that determines the type of contagion, e.g., simple, deterministic complex, or noisy complex; see SI Figure S13 for a block diagram of state transitions. Activation functions for different models of contagion are characterized by different parameters, e.g., the independent transmission probabilities ($\beta$) for the simple contagion activation functions in Figure \ref{fig:simple_activation_functions}, or the threshold value ($\theta$) and sub-threshold adoption probability ($q$) for the complex contagion activation functions in Figure \ref{fig:complex_activation_functions}.  In addition to the transitions that we investigate in our theoretical analysis (e.g., reversion from infected to susceptible with probability $\delta$), our simulations include transition probabilities between active and inactive infected states, to model situations that adopter agents transition into an ``inactive infected'' state where they do not influence their neighbors but remain adopters.\\

\noindent\textbf{Empirical networks data.} The empirical network data for our simulation studies are derived from publicly available data \cite{chami2017social,cai2015social,banerjee2013diffusion,traud2012social}. The Cai et al. \cite{cai2015social} data is comprised of $175$ social networks of Chinese farm villages that are collected in the study of farmers being encouraged to sign up for a weather insurance product. The friendship and health advice network data are collected by Chami et al. \cite{chami2017social} from $17$ rural villages in Uganda. The Banerjee et al. \cite{banerjee2013diffusion} data contains the interconnection data for multi-dimensional social relations in $77$ villages in southern India. Traud et al. \cite{traud2012social} data contains the Facebook friendship networks at U.S. colleges and universities; we use the $40$ smallest networks, for which such simulations are more computationally practical. The village networks in the first three sets have as few as tens of nodes but have typically hundreds of nodes. A typical Facebook college network has thousands of nodes. SI Table S1 summarizes the statistics for each set of networks.
{\small
\section*{Data availability}
The simulations on empirical networks use publicly available data \cite{chami2017social,cai2015social,banerjee2013diffusion,traud2012social}.

\section*{Code availability}
Code for reported simulations can be accessed from \href{https://github.com/aminrahimian/social-contagion/wiki}{https://github.com/aminrahimian/social-contagion/wiki}.

\section*{Acknowledgements}{Authors are listed alphabetically. Mossel was partially supported by NSF grant CCF 1665252, DOD ONR grant N00014-17-1-2598, and NSF grant DMS-1737944. Rahimian acknowledges support from Pitt Momentum Funds and a Pitt Cyber Accelerator grant. This research was supported in part by the University of Pittsburgh Center for Research Computing, RRID:SCR\_022735, through the resources provided. Specifically, this work used the H2P cluster, which is supported by NSF award number OAC-2117681. During his postdoctoral work at MIT, Rahimian was supported by an Amazon Research Award to Eckles. We thank Carlos Hurtado, Yixuan Long and Clinton S. Reid for research assistance. We thank Sinan Aral, Stephen Morris, and David G. Rand for helpful comments. We also thank James Moody, two other anonymous referees, and editors at Nature Human Behaviour for their helpful feedback in revising the manuscript.}

\section*{Potential competing interests}
Meta (which operates Facebook) has sponsored a conference co-organized by Eckles and has funded some of his other research. Rahimian has served on the advisory committee of a vaccine confidence fund created by Meta and Merck, and some of his research has been also funded by Meta.
}
\bibliographystyle{Science}

\bibliography{ref.bib}{References}


\includepdf[pages={1-}]{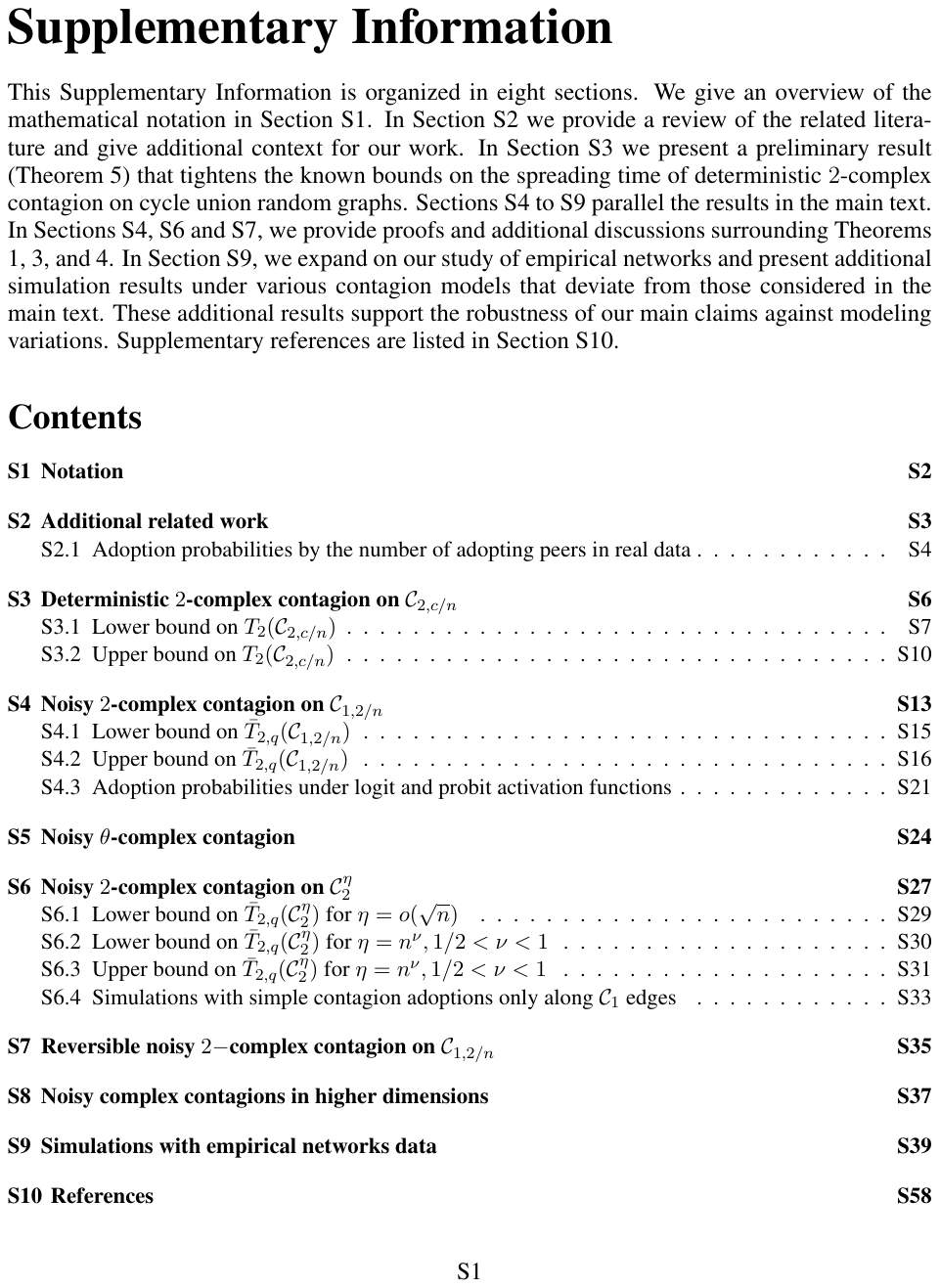}

\end{document}